\documentclass[prl, twocolumn, superscriptaddress]{revtex4-1}
\usepackage{amsmath,amsbsy, amssymb}
\usepackage[dvips]{graphicx}
\usepackage{ulem}
\usepackage{xcolor}
\def\half{{\textstyle \frac{1}{2}}}

\def\fourfifths{{\textstyle \frac{4}{5}}}
\def\threequarters{{\textstyle \frac{3}{4}}}

\def\Tr#1{{\rm Tr}\left( #1 \right)}
\def\Det#1{{\rm Det}\!\left( #1 \right)}
\begin{document}
\title{Fluid-driven fingering instability of a confined elastic meniscus}
\author{J.\ S.\ Biggins}
\affiliation{Cavendish Laboratory, 19 JJ Thomson Ave, Cambridge University, Cambridge, United Kingdom}
\author{Z.\ Wei}
\affiliation{School of Engineering and Applied Sciences, Harvard University, Cambridge, MA 02138, USA}
\author{L.\ Mahadevan}
\affiliation{School of Engineering and Applied Sciences, Harvard University, Cambridge, MA 02138, USA}
\affiliation{Department of Physics, Harvard University, Cambridge, MA 02138, USA}
\date{\today}

\begin{abstract}
When a fluid is pumped into a cavity in a confined elastic  layer, at a critical pressure, destabilizing fingers of fluid invade the elastic solid along its meniscus \cite{Bouchaudfinger}. These fingers occur without fracture or loss of adhesion and are reversible, disappearing when the pressure is decreased.  We develop an asymptotic theory of pressurized highly elastic layers trapped between rigid bodies to explain these observations, with predictions for the critical fluid pressure for fingering, and the finger wavelength. We also show that the theory links this fluid-driven fingering with a similar transition driven instead by transverse stretching of the elastic layer. We further verify these predictions by using finite-element simulations on the two systems which show that, in both cases, the fingering transition is first-order (sudden) and hence has a region of bistability. Our predictions are in good agreement with recent observations of this elastic analog of the classical Saffman-Taylor interfacial instability in hydrodynamics.
\end{abstract} 

\maketitle
\begin{figure}[h]
\includegraphics[width=\columnwidth]{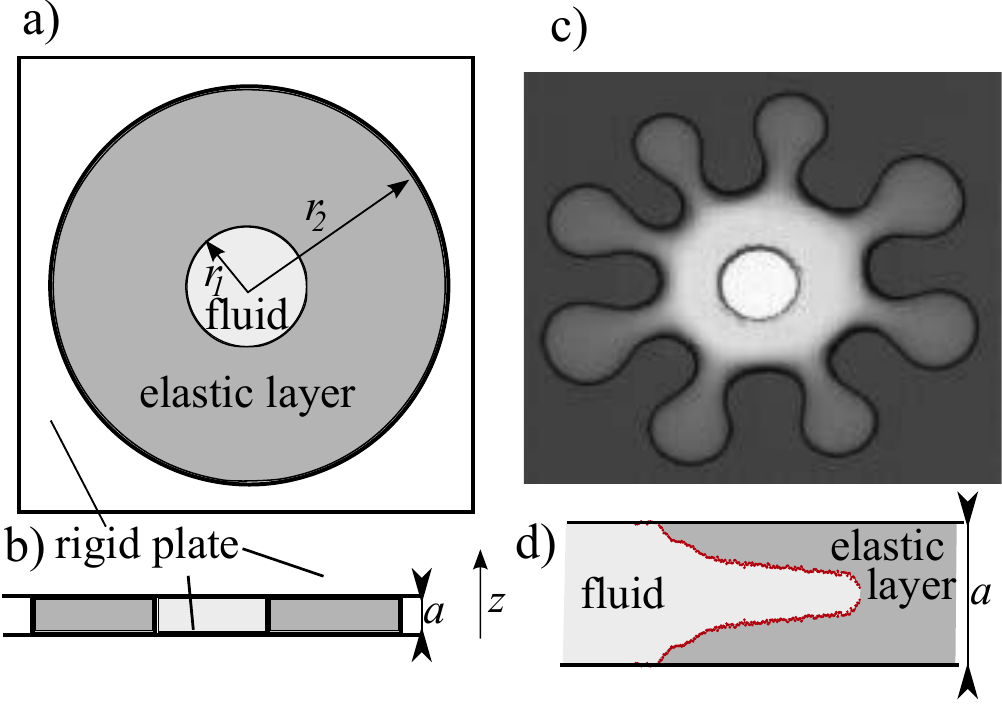}
\caption{(a) Top-view of the experimental setup used to study the elastic analog of the Saffman-Taylor finger \cite{Bouchaudfinger}: two rigid plates confine a thin elastic layer with a central cavity containing fluid whose volume is increased by injecting fluid (from above). (b) Cross-section showing the thickness of the elastic layer. (c) Experimentally obtained fingering pattern \cite{Bouchaudfinger}. The central hole corresponds to the original cavity size, while the varying gray scale is a consequence of the  elastic meniscus deforming substantially without loss of adhesion to the plates. (d) Experimental cross-section of a finger \cite{Bouchaudfinger} showing maintained adhesion. }\label{bouchaudexpfigure}
\end{figure}

In continuum mechanics, fingering instabilities are usually associated with interfacial  flows in porous media, or its analog, flow in a Hele-Shaw cell. Indeed, the prototypical interfacial instability is the celebrated Saffman-Taylor fingering, wherein a viscous fluid is confined between two plates and, when a less viscous fluid is pumped in, their interface becomes unstable and the less visous fluid invades in finger like protrusions\cite{Saffman1958}.   Recently the elastic analog of the Saffman-Taylor experiment was explored by pumping a fluid into a cavity in a confined elastic layer\cite{Bouchaudfinger}. This causes the cavity to first dilate laterally without any loss of adhesion between the elastic solid and the confining plates. At a critical pressure,  fingers of fluid invade the elastic layer, as seen in Fig.\ \ref{bouchaudexpfigure}, just like classical Saffman-Taylor fingers, with the viscous fluid replaced by a highly elastic solid. Related fingering transitions have also been reported in thin confined layers of soft elastic solids that sit betwixt nominally rigid bodies which are pulled apart. In one case, peeling causes adhesion between the layer and  body to fail, and  finger-like undulations appear along the resulting contact line \cite{ghatak2000meniscus,adda2006crack}. In another case, adhesion is maintained and finger-like invaginations  appear at the perimeter of  the elastic layer \cite{shull2000fingering,bigginsfingering} when the rigid bodies are pulled apart. Both these transitions have been compared to Saffman-Taylor fingering but, since there is no analogue of the invading fluid, the analogy is somewhat superficial.  Here we provide a  theoretical understanding of the elastic Saffman-Taylor fingering instability, and provide a unifying treatment of fingering in thin elastic layers produced either by lateral fluid invasion or by transverse layer dilation with maintained adhesion, showing that they lead to identical patterns. 

We begin with scaling estimates for fluid-driven elastic fingering in a thin incompressible neo-Hookean annular layer (fig.\ \ref{bouchaudexpfigure}a-b)  adhered to rigid plates at $z=\pm a/2$  and with in-plane extent $r_1<r<r_2$  and shear modulus $\mu$. Since adhesion is maintained, an in-plane displacement $u$, applied mid-way between the plates, will generate strains $\gamma\sim u/a$ localized in-plane by an elastic screening length of $\mathcal{O}(a)$. A fluid (pressure $P_f$) pumped into the cavity  will induce such a displacement radially on the inner circumference, increasing its volume by $\delta V_f\sim 2 \pi r_1 u a$. Since the layer is incompressible, this $u$ cannot be screened but  decays radially as $u(r)\sim u r_1/r$. Equating the layer's  elastic energy, $E\sim a \int_{r_1}^{r_2} \half \mu  \gamma^2 2 \pi r dr \sim \mu  u^2 r_1^2 \log(r_2/r_1)/a$, and the fluid's work,  $P_f \delta V_f$,  we predict $u\sim  (P_f/\mu) a^2/(r_1\log(r_2/r_1))$. The only non-linearities available to drive fingering are large-strain geometric ones,   important when $\gamma\gtrapprox1$, requiring a threshold $P_f\sim\mu (r_1/a)\log(r_2/r_1)$. Sinusoidal perturbations on the interface will be screened,  so the finger wavelength will  scale as $a$.

To verify and improve these estimates, we  build a minimal 2-d theory, taking advantage of the geometric scale separation induced by confinement. Consider a point with position vector $\mathbf{R}=\mathbf{r}+z \mathbf{\hat{z}}$, and displacement $\mathbf{V}(\mathbf{R})=\mathbf{u}(\mathbf{R})+v_{\perp}(\mathbf{R})\mathbf{\hat{z}}$, where $\mathbf{r}$ and $\mathbf{u}$ are in-plane vectors, and $\mathbf{\hat{z}}$ is the layer normal. Expanding  $\mathbf{V}(\mathbf{R})$ to second order in $z$, imposing symmetry around  $z=0$ and requiring  $\mathbf{V}(\mathbf{R})=0$ at $z=\pm a/2$ we get an approximate form for the displacement,
\begin{align}
\mathbf{V}(\mathbf{R})=(1-2 z/a)(1+2 z /a) \mathbf{u}(\mathbf{r}).
\end{align}
Soft incompressible solids are well modeled  by the neo-Hookean energy density $\half \mu (\Tr{F\cdot F^{T}}-3)$, where $F_{\alpha \beta}=\delta_{\alpha \beta}+\partial_\beta \mathbf{V}_{\alpha}$ is the deformation gradient, and incompressibility requires  $\Det{F}=1$. Implementing incompressibility in a depth-averaged sense for thin layers, we define our  2-d energy density by
\begin{equation}
L= \int_{-a/2}^{a/2}\half\mu\left(\Tr{F\cdot F^T}-3\right)-P(\Det{F}-1) \mathrm{d}z.
\end{equation}
The quadratic form for $\mathbf{V}$ means $F=I+(1-4 z^2/a)\nabla \mathbf{u}(\mathbf{r})-8 z/a^2 \mathbf{u}(\mathbf{r}) \mathbf{\hat{z}}+\mathbf{z}\mathbf{z}$, where $I$ and $\nabla$ are the in-plane identity and gradient.  Conducting the thickness ($z$) integral gives
\begin{align}
&L(\mathbf{u},P)=\\
&\frac{5 a }{6} \left(\frac{1}{2}\mu(\Tr{G\cdot G^T}-2)+ \frac{16}{5}\mu \frac{\mathbf{u}\cdot \mathbf{u}}{a^2}-P(\Det{G}-1)\right)\notag
\end{align}
where  $G=I+\fourfifths \nabla \mathbf{u}$ is an  effective 2-d deformation gradient, and $P$ is a 2-d pressure field.  Minimizing the total elastic energy $E=\int L \mathrm{d}A$ over  $\mathbf{u}$ and $P$  leads to the Euler-Lagrange equations
\begin{align}
&\frac{8\mu  }{ a^2 }\mathbf{u}=\frac{4\mu  }{5}\nabla^2 \mathbf{u}- \Det{G}G^{-T}\cdot \nabla P\label{E_L1}\\
&\Det{G}=1.\label{E_L2}
\end{align}
To derive the associated boundary conditions, we imagine a small additional displacement $\delta\mathbf{u}$ that gives rise to a change in $E$ arising at the boundary $\delta E=\frac{2 a }{3}\oint\delta \mathbf{u}\cdot\left(\mu G-P\Det{G}G^{-T}\right)\cdot\mathbf{\hat{n}}.\mathrm{d}s$, where $\mathbf{\hat{n}}$ is the boundary's outward  normal. At a free boundary $\delta E$ would vanish. At an interface with fluid at pressure $P_f$ we  must add the virtual work term $-P_f V_f$ ($V_f$ is the fluid volume) to $E$, generating an additional boundary term $-P_f \delta V_f$. A  small patch of  boundary at height $z$, thickness $\mathrm{d}z$ and in-plane extent $\mathrm{d}s$ has initial vector area $\mathrm{d}\mathbf{A}=\mathrm{d}z\mathrm{d}s\mathbf{\hat{n}}$. After deformation, this  becomes $\Det{F}F^{-T}\cdot\mathrm{d}\mathbf{A}$. An incremental displacement $\delta \mathbf{u}$   displaces the patch by $(1-4 z^2/a^2)\delta \mathbf{u}$ and hence changes the fluid volume by $-(1-4 z^2/a^2)\delta \mathbf{u}\cdot \Det{F}F^{-T}\cdot\mathrm{d}\mathbf{A}$. Integrating this over the boundary gives $\delta V_f=-\oint \delta \mathbf{u}\cdot\int_{-a/2}^{a/2}(1-4 z^2/a^2)\Det{F}F^{-T}\mathrm{d}z\cdot\mathbf{\hat{n}}.\mathrm{d}s$. Conducting the  $z$ integral then gives $\delta V_f=-\frac{2 a}{3}\oint \delta \mathbf{u}\cdot\Det{G}G^{-T}\cdot\mathbf{\hat{n}}.\mathrm{d}s$ and hence the appropriate boundary conditions are
 \begin{equation}
(\mu G+(P_f-P)\Det{G}G^{-T})\cdot{\mathbf{\hat{n}}}=0,\label{bc}
\end{equation}
which, with eqns (\ref{E_L1}-\ref{E_L2}), specify the problem.

We first  solve these equations for fingering in a simple Cartesian geometry, considering an elastic layer in an infinite strip with $0<y<l$ and $-\infty<x<\infty$, an invading fluid at pressure $P_f$ for $y<0$ and a vacuum for $y>l$. We expect  fingering of the   $y=0$ boundary at a critical $P_f$, so we write the fields as a translationally invariant base-state plus a small  perturbation:
\begin{align}
\mathbf{u}&=Y_1(y)\mathbf{\hat{y}}+\epsilon \mathbf{u_2(x,y)},\hspace{1.5em}P=P_1(y)+\epsilon P_2(x,y).
\end{align}
Substituting these  into eqns  (\ref{E_L1}-\ref{E_L2}) and setting  $\epsilon=0$, we  see that  $Y_1$ is a constant and $P_1$ is linear in $y$. Applying  eqn.\ (\ref{bc}) at $y=0$ and at $y=l$ (where $P_f=0$) then yields:
\begin{equation}
Y_1(y)=a^2 P_f/(8 l \mu),\hspace{2em}P_1(y)=\mu+P_f-P_f y/l\label{stripbase}.
\end{equation}
Expanding eqns (\ref{E_L1}-\ref{bc}) to linear order in $\epsilon$ around this state gives us an eigenvalue problem for the base state's stability
\begin{align}
&\frac{8 \mu}{ a^2 }\mathbf{u_2}=\frac{4\mu  }{5}\nabla^2 \mathbf{u_2}-  \nabla P_2+\frac{4}{5}\nabla P_1\cdot (\nabla \mathbf{u_2})\label{pert1}\\
&\nabla\cdot\mathbf{u_2}=0,\\
&\left(\fourfifths \mu \nabla \mathbf{u_2}-P_2+ \fourfifths \mu (\nabla \mathbf{u_2})^T\right)\cdot \mathbf{\hat{n}}=0\label{pert3}.
\end{align}
Assuming explicit oscillatory  perturbative fields,  $P_2=P_2(y)  \cos(k x)$,  $\mathbf{u_2}=Y_2(y)\cos(k x)\mathbf{\hat{y}}+X_2(y)\sin(k x)\mathbf{\hat{x}}$, we solve these equations and see that, provided $l\gg a$, the boundary destabilizes when
\begin{equation}
P_f=\frac{2 \mu l}{5 a}\hspace{0.1em}\frac{a^2 k^2 \left(a k \left(a k-\sqrt{a^2 k^2+10}\right)+10\right)+25}{a k}.
\end{equation}
Minimizing this  threshold over $k$, we see that fluid-driven fingering of a rectilinear elastic meniscus  occurs with wavelength and pressure
\begin{equation}
\lambda \approx 2.75... a \hspace{3em}P_f \approx 10.1... l \mu/a.\label{pthresh}
\end{equation}

We next consider the experimental circular geometry  \cite{Bouchaudfinger}. A naive extrapolation of our Cartesian stability analysis result to the circular case by taking $l\sim r_2-r_1$ would predict threshold pressures far beyond those observed because the Cartesian base-state is 1-D whereas in the circular one is 2-d, with different qualitative forms for the decay of the elastic fields.  Assuming an annular   elastic layer occupying the region $r_1<r<r_2$, $-\pi<\theta<\pi$ with a fluid at pressure $P_f$ in the cavity $r<r_1$ and a vacuum for $r>r_2$ allows us to write the displacement and pressure fields as
\begin{align}
\mathbf{u}&=R_1(r)\mathbf{\hat{r}}+\epsilon (R_2(r)\cos(n \theta) \mathbf{\hat{r}}+\Theta_2(r)\sin(n \theta) \boldsymbol{\hat{\theta}})\\
P&=P_1(r)+\epsilon P_2(r) \cos(n \theta).
\end{align}
Substituting these expressions into (\ref{E_L1}-\ref{E_L2}),  then setting $\epsilon=0$, allows us to solve (\ref{E_L2}) for $R_1$,
\begin{equation}
R_1(r)=\frac{5r }{4}\left(\sqrt{1+\left(\frac{c_4}{r}\right)^2}-1\right),\label{fullr1}
\end{equation}
where the integration constant $c_4$   parameterizes the inner boundary's displacement. We can solve for $P_1$ analytically  then solve the perturbative equations (\ref{pert1}-\ref{pert3}) numerically to find the fingering threshold and mode without further approximation (see SI) but the algebra is cumbersome. However, the expressions simplify in the limit of thin layers, $a\ll r_1$, a case of much interest. As in the Cartesian geometry, we expect an instability when $R_1(r_1)\sim a$,  when strains become geometrically large. Such displacements require $c_4\sim \sqrt{r_1 a}\ll r_1$, so  $R_1$ can be replaced by its first order expansion $R_1(r)=5 c_4^2/(8 r)$. Furthermore, $R_1'(r)\sim c_4^2/r^2$ is negligibly small so we can  neglect gradients of $\mathbf{u}$,  setting $G=I$.  This reduces eqn.\ (\ref{E_L1}) to $\frac{8\mu  }{ a^2 }R_1(r)=-  P_1'(r)$, which on integration yields  $P_1 \sim \log(r)$. Similarly applying the boundary conditions (\ref{bc}) allows us to determine $R_1, P_1$ as:
\begin{equation}
R_1(r)=\frac{a^2 P_f}{8 \mu  r \log \left(r_2/r_1\right)},\hspace{0.03em}
P_1(r)=\mu +\frac{P_f \log \left(r/r_2\right)}{\log \left(r_1/r_2\right)}.
\end{equation}
Both these fields only vary on length-scales comparable to $r_1$, so in a region around the inner boundary with $r_1 \ll r \ll a$ they are well described by their Taylor expansions around $r_1$ given by:
\begin{align}
R_1\hspace{-0.13em}=\hspace{-0.13em}\frac{a^2 P_f}{8 \mu  r_1 \log \left(r_2/r_1\right)}, \hspace{0.2em} P_1\hspace{-0.2em}=\hspace{-0.1em}\mu \hspace{-0.1em}+\hspace{-0.13em}P_f+\hspace{-0.13em}\frac{P_f (r-r_1)}{r_1 \log \left(r_1/r_2\right)}\hspace{-0.2em}.\label{pdrivenbasetaylor}
\end{align}
Identifying $(r-r_1)\to y$ and $r_1\log\left(r_2/r_1\right)\to l$,  these results match the base state for the rectilinear case  (eqn.\ (\ref{stripbase})).  Fingering only occurs within a characteristic distance $a$ from the boundary where base states match, so the instability will proceed in the same way with mode-number ($n=2\pi r_1/\lambda$) and threshold
\begin{equation}
n\approx 2.28 r_1/a, \hspace{2em}P_f \approx 10.1 \mu (r_1/a) \log\left(r_2/r_1\right).\label{assymtopes}
\end{equation}
This pressure diverges logarithmically as $r_2\to \infty$ so fingering will occur in a pressurized cavity in an almost infinite layer, but not in a   wide rectilinear strip.

\begin{figure}
\includegraphics[width = \columnwidth]{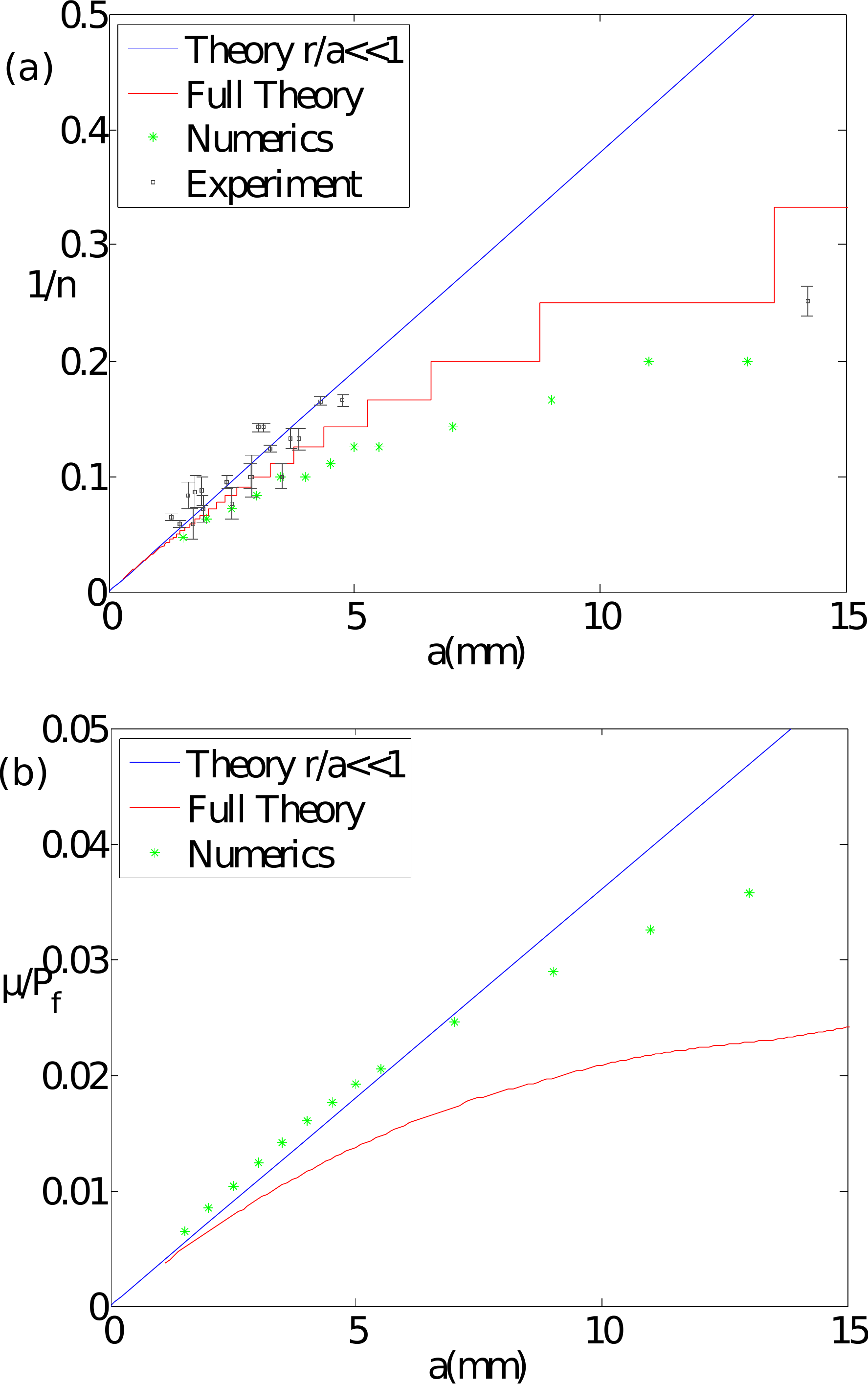}
\caption{A thin circular elastic layer with thickness $a$, shear modulus $\mu$ and radius $r_2=125\mathrm{mm}$ has a central cavity of radius $r_1=11.5\mathrm{mm}$ filled with a fluid at pressure $P_f$. Above a threshold $P_f$, fingers of fluid invades the layer. We show the inverse number of fingers $1/n$ (left) and the inverse scaled threshold pressure $\mu/P_f$ (right) as a function of the layer thickness. The plots compare the predictions of the full 2D theory based on eqn.\ (\ref{fullr1}) (red lines), the asymptotic results for $a/r_1\ll1$ given in eqn.\ (\ref{assymtopes}) (blue lines), full finite-element results and, in the left plot, experimental results \cite{Bouchaudfinger}.}
\label{numsol}
\end{figure}

In Fig. \ref{numsol}, we compare these predictions with experiments  \cite{Bouchaudfinger} and finite element simulations carried out using a commercial package ABAQUS and see that the three agree well for very thin layers. Our data extends to layers with $a/r_1\gtrapprox 1$ which are not thin; unsurprisingly, here the depth-averaged asymptotic theory predicts too few fingers and too high pressures.  A better approximation can be obtained by returning to the full expression for $R_1(r)$  (eqn.\ (\ref{fullr1})) and continuing the derivation without assuming  $a\ll r_1$ (see SI), and are also shown in Fig.\ \ref{numsol}.  The theory is still  depth-averaged so it does not capture the full behavior of thick layers, but it captures the qualitative nature of the non-linear deviations.

As alluded to in our introduction, fingering of a confined elastic layer can also be driven by  transverse displacement \cite{shull2000fingering, bigginsfingering}.  Layer incompressibility implies that pulling the plates apart causes the meniscus to be inwardly displaced and, at a critical separation, fingers form in a manner reminiscent of  fig.\ \ref{bouchaudexpfigure}. The similarity arises despite the difference in the origin of the base-states because both add volume to an incompressible layer, resulting in long-ranged displacements that only vary on in-plane length-scales. In the boundary region of characteristic width $a$ where fingering occurs, both base states are essentially constant inward displacements, and finger identically.  We now show how our theory makes this connection concrete.
If the invading fluid is removed ($P_f=0$) and instead the rigid plates are separated to $z=\pm (a+\Delta z) /2$ we must modify $\mathbf{V}(\mathbf{R})$ to 
\begin{equation}
\mathbf{V}(\mathbf{R})=(1-2 z/a)(1+2 z /a) \mathbf{u}(\mathbf{r})+z \mathbf{\hat{z}}\Delta z/a.
\end{equation}
Since separation adds volume to the whole layer area, while the inward displacement only does so at the boundary, for thin wide layers, the $\Delta z$ required for displacement comparable to $a$ will be small. Assuming $\Delta z/a\ll1$, the above $\mathbf{V}$ leads to the equations of equilibrium \cite{bigginsfingering}
\begin{align}
&\frac{8\mu  }{ a^2 }\mathbf{u}=\frac{4\mu  }{5}\nabla^2 \mathbf{u}- \Det{G}G^{-T}\cdot \nabla P\label{peqnd},\\
&\Det{G}=1-6 \Delta z/(5a),\label{deteqn}\\
&(\mu G-P\Det{G}G^{-T})\cdot{\mathbf{\hat{n}}}=0\label{bceqn},
\end{align}
identical to  the pressure driven case, except the driving term has changed from $P_f$ in  the boundary condition to  $6 \Delta z/(5a)$  in  eqn.\ (\ref{deteqn}). 

In the Cartesian strip geometry, we can solve eqn.\ (\ref{deteqn}) for the translationally invariant displacement $Y_1(y)=\threequarters a (l \Delta z /a^2) (1 - 2 (y/l))$, which is symmetric about $y=l/2$ and hence substantially different to the pressure driven case.  However, since it only varies over distances comparable to $l$, in a region of width comparable to $a$ around the $y=0$ boundary, it is is essentially constant, $Y_1(0)=\threequarters l \Delta z/a$.   Substituting this constant into eqns.\ (\ref{peqnd}) and (\ref{bceqn}), we see that, in the same boundary region, the pressure is given by $P=\mu-6 \mu l y \Delta z/a^3$. Thus, identifying  $\Delta z\to a^3 P_f/(6 l^2 \mu)$, in this boundary region the separation-driven fields match the pressure driven ones (eqn.\ (\ref{stripbase})), up to an offset $P_f$ in the pressure.

We next consider the stability of these base states by considering small perturbations, $P=P_1(y)+\epsilon P_2(x,y)$ and $\mathbf{u}=\mathbf{u}_1(y)+\epsilon \mathbf{u}_2(x,y)$,  localized to the $y=0$ boundary. If we expand eqns.\ (\ref{peqnd}-\ref{bceqn}) to first order in $\epsilon$, this is analogous deriving eqns (\ref{pert1}-\ref{pert3}). The only two differences are the offset in the base pressures by $P_f$ , which simply cancels the offset by $P_f$ between the two boundary conditions, and the $6 \Delta z/(5a)$ term in eqn.\ (\ref{deteqn}) which, in the thin layer limit, is negligibly small. Thus the stability of a thin layer is also governed by eqns.\  (\ref{pert1}-\ref{pert3}), and the instability proceeds in the same way, with threshold $\Delta z\approx 1.68a^2/l$. The same reasoning applies even with large perturbations, so the full non-linear finger development is identical. 

In the annular geometry, we solve eqn.\ (\ref{deteqn}) for the  base state to get $R_1(r)=\frac{5 r}{4} \left(\sqrt{ 1-\frac{6 \Delta z}{5 a}+\left(\frac{c_4}{r}\right)^2}-1\right)$. As in the pressure-driven case, for thin layers with $a/r_1 \ll 1$, we may expand the root in the previous expression to get $R_1(r)=\frac{5 c_4^2}{8  r}-\frac{3 \Delta z r}{4 a}$. Solving eqns (\ref{peqnd}-\ref{bceqn}) for the full base state then yields
\begin{align}
&P_1(r)=\mu\hspace{1em}+\\&\frac{3 \Delta z \mu}{a^3 \log \left(r_1/r_2\right)}  \left(r^2 \log \left(\frac{r_1}{r_2}\right)+r_1^2 \log \left(\frac{r_2}{r}\right)+r_2^2 \log \left(\frac{r}{r_1}\right)\right)\notag\\
&R_1(r)=\frac{3 \Delta z \left(r_1^2-r_2^2\right)}{8 a r \log \left(r_1/r_2\right)}-\frac{3 \Delta z r}{4 a}.
\end{align}
These fields vary on length-scales comparable to $r_1\gg a$, so in a region around the inner boundary with characteristic width $a$ they are well approximated by their Taylor series around $r_1$. Identifying
\begin{equation}
\Delta z\to\frac{P_f}{3 \mu }\frac{a^3}{ 2 r_1^2 \log \left(r_1/r_2\right)-r_1^2+r_2^2}\label{dzeqiv},
\end{equation}
we see that the equivalent series  differ from those in the pressure driven case (eqn.\ \ref{pdrivenbasetaylor}) by the same offset of $P_f$ to $P_1$ as in the Cartesian strip case. Thus, as before, the base states differ on long length scales but match around the inner boundary, and  are susceptible to exactly the same fingering instability. Substituting the threshold pressure  into the above expression for $\Delta z$, we find the threshold separation for fingering which, when $r_2\gg r_1$, reduces to  $\frac{\Delta z}{a} \approx 3.37 \frac{a}{r_2}\frac{r_1}{r_2}\log{\left(r_2/r_1\right)}$, and is indeed  small.

\begin{figure}\centering
\includegraphics[width=\columnwidth]{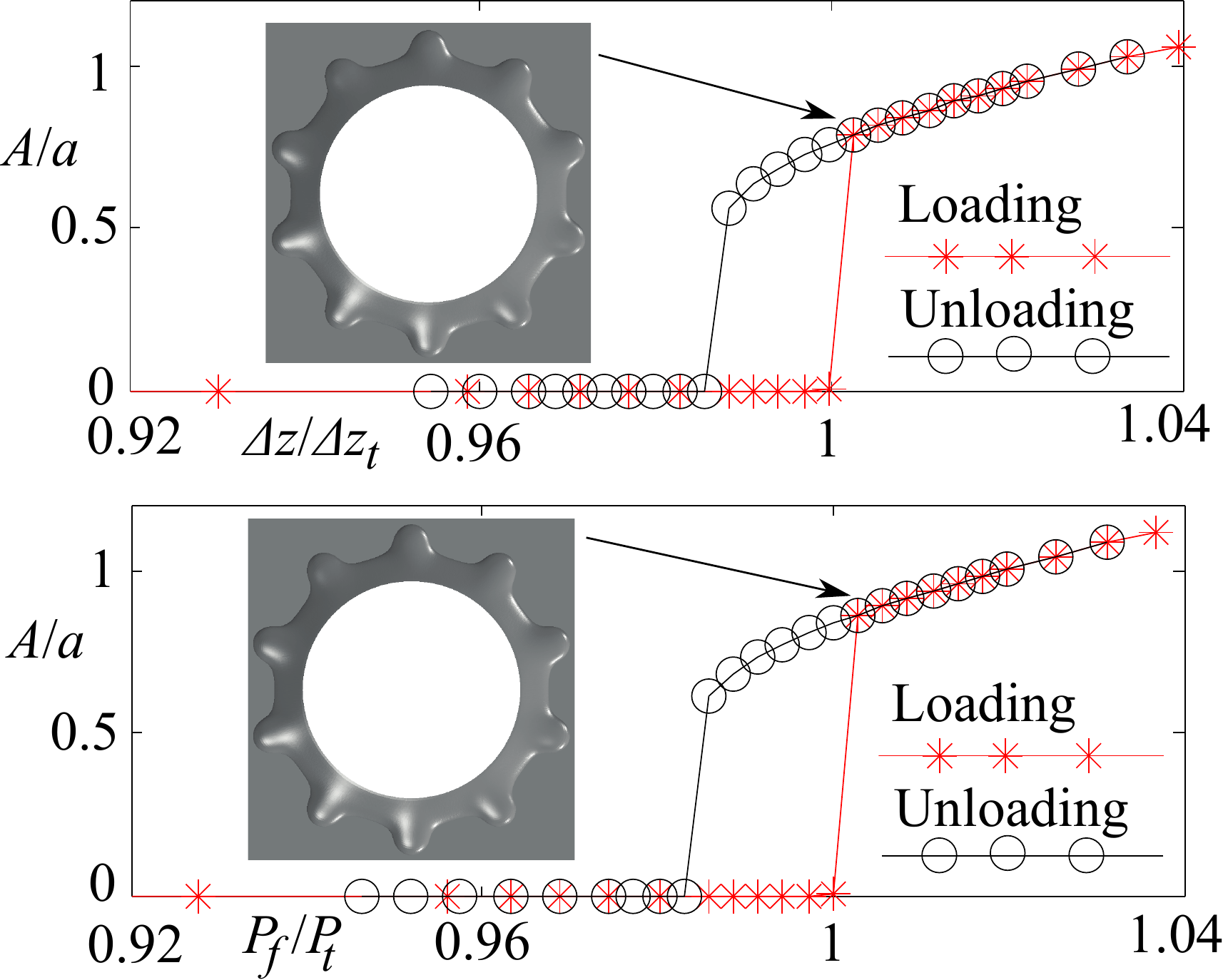}
\caption{Finite element hysteresis loops showing finger amplitude $A$ for displacement   (top) and fluid pressure (bottom) driven fingering, using  $a=3.5m$m $r_1=11.5m$m and $r_2=125m$m. Both show a first order transition to very similar fingered states, (see  insets) at threshold separation $\Delta z_t=0.02 a$ and pressure $P_t=69.4\mu$ respectively. The dimensionless threshold ratio $(P_t/\mu)/(\Delta z_t/a)=3510\pm10$ is close to theoretical estimate of 3640 from eqn.\ (\ref{dzeqiv}).}\label{numtests}
\end{figure}

We confirm this equivalence between fluid and displacement-driven fingering via  ABAQUS finite element simulations. Fig.\ \ref{numtests} shows the hysteresis loops and fingering patterns for the two cases. Despite the layers being only modestly thin ($r_1/a\sim0.3$), the loops are  very similar. The fingering transition is sub-critical in both cases, and hence both systems exhibit bistability.  

Our study highlights the geometrical similarity and the essential physical differences between elastic and viscous fingering. Elastic fingering is governed by an equilibrium first-order transition whilst viscous fingering is a rate-dependent dynamic process with a continuous transition driven by a competition between surface tension ($\gamma$) and viscous shear/ pressure gradients. Surface tension will become important in elastic fingering if the  layer thickness becomes comparable to the elastocapillary length scale  $\gamma/\mu$. Furthermore, one could interpolate between the elastic and viscous limits using viscoelastic materials,  unifying a broad range of invasive fingering phenomena. These may be relevant to many phenomena in adhesion science/engineering and perhaps even biological morphogenetic processes where branching and fingering abound.

\begin{acknowledgments}
We thank Elisabeth Bouchaud and Baudouin Saintyves for introducing us to this experiment, for useful discussions and Fig. 1c,d.  We also thank Trinity Hall, Cambridge and the 1851 Royal Commission (JSB), the Harvard NSF-MRSEC DMR0820484 (ZW,LM) and the MacArthur Foundation (LM) for partial support. \end{acknowledgments}

\onecolumngrid
\newpage

\begingroup  
  \centering
  \LARGE Supplementary Information for "Fluid driven fingering instability of a confined elastic meniscus"\\ [1em]
   \endgroup
\begingroup
\centering
J.\ S.\ Biggins, Z.\ Wei \& L. Mahadevan

\endgroup

Here, we provide details of our calculations that were algebraically too tedious to be presented in the main text.

As in our main manuscript, we consider an annular neo-Hookean elastic layer with in-plane extent $r_1<r<r_2$ and thickness $a$ that is bound to rigid plates at $z=\pm a/2$. A fluid with pressure $P_f$ is pumped into the central cavity ($r<r_1$), while for $r>r_2$ there is a vacuum. In our manuscript we show that the deformations of such a layer will be governed by the 2-D bulk equations 
\begin{align}
&\frac{8\mu  }{ a^2 }\mathbf{u}=\frac{4\mu  }{5}\nabla^2 \mathbf{u}- \Det{G}G^{-T}\cdot \nabla P\label{SI-E_L1}\\
&\Det{G}=1,\label{SI-E_L2}
\end{align}
and the boundary condition
 \begin{equation}
(\mu G+(P_f-P)\Det{G}G^{-T})\cdot{\mathbf{\hat{n}}}=0,\label{SI-bc}
\end{equation}
where $P$ is a 2-D pressure field, $P_f$ is the fluid pressure on the boundary, $\mathbf{\hat{n}}$ is the outward normal on the boundary, $\mathbf{u}$ is a 2-D in-plane displacement, $\mu$ is the shear modulus, and $G=I+\fourfifths \nabla \mathbf{u}$ is an effective deformation gradient. These equations are eqns (\ref{E_L1}-\ref{bc}) in our main manuscript. Since we are working in a $(r, \theta)$ circular polar coordinate system we first recall the forms of the gradient operators in these equations, using commas to denote partial derivatives, 
\begin{equation}
\nabla P=\left( \begin{array}{c}
P,_{r}\\
\frac{P,_{\theta}}{r}
\end{array}\right)\hspace{2em}\nabla \mathbf{u}=\left( \begin{array}{cc}
u_r,_r&\frac{u_r,_\theta-u_\theta}{r}\\
u_{\theta},_{r}&\frac{u_\theta,_\theta+u_r}{r}
\end{array}\right)\hspace{2em}\nabla^2 \mathbf{u}=\left( \begin{array}{c}
u_r,_{rr}+\frac{u_r,_{\theta\theta}}{r^2}+\frac{u_r,_r}{r}-\frac{2 u_\theta,_\theta}{r^2}-\frac{u_r}{r^2}\\
u_\theta,_{rr}+\frac{u_\theta,_{\theta\theta}}{r^2}+\frac{u_\theta,_r}{r}+\frac{2 u_r,_\theta}{r^2}-\frac{u_\theta}{r^2}
\end{array}\right).
\end{equation}

Before considering the interfacial stability of the inner boundary, we first consider the layer's initial azimuthally symmetric response
\begin{equation}
\mathbf{u}=R_1(r)\mathbf{\hat{r}}\hspace{3em}P=P_1(r).
\end{equation}	
We then have 
\begin{equation}
G_1=\left( \begin{array}{cc}
1+\fourfifths R_1'(r)&0\\
0&1+\fourfifths R_1(r)/r
\end{array}\right),\hspace{3em}
\Det{G_1}G_1^{-T}=\left( \begin{array}{cc}
1+\fourfifths R_1(r)/r&0\\
0&1+\fourfifths R_1'(r)
\end{array}\right)\label{SI-G1eqn}
\end{equation}so eqn.\ \ref{SI-E_L2} has solution
\begin{equation}
R_1(r)=\frac{5r }{4}\left(\sqrt{1+\left(\frac{c_4}{r}\right)^2}-1\right),\label{SI-fullr1}
\end{equation}
where $c_4$ is a constant of integration. The $\mathbf{\hat{r}}$ component of eqn.\ \ref{SI-E_L1} then reduces to
\begin{equation}
\frac{8 \mu}{a^2}R_1(r)=\frac{4\mu}{5}\left(\frac{r R_1'(r)-R_1(r)}{r^2}+R_1(r)\right)+\left(1+\frac{4  R_1(r)}{5 r}\right) P_1'(r),
\end{equation}
Substituting in eqn.\ (\ref{SI-fullr1}) for $R_1$ and simplifying, this equation reduces to
\begin{equation}
a^2 r \left(c_4^2+r^2\right)^2 P_1'(r)+\mu  \left(a^2 c_4^4+10 r^2 \left(c_4^2+r^2\right) \left(-r \sqrt{c_4^2+r^2}+c_4^2+r^2\right)\right)=0
\end{equation}
and solving this equation, the base-state pressure is
\begin{equation}
P=P_0-\mu  \left(\frac{5 r^2}{a^2} \left(1-\sqrt{\frac{c_4^2}{r^2}+1}\right)+\frac{5 c_4^2}{a^2} \log \left(\sqrt{c_4^2+r^2}+r\right)+\frac{c_4^2}{2 \left(c_4^2+r^2\right)}-\log \left(\sqrt{\left(\frac{c_4}{r}\right)^2+1}\right)\right),
\end{equation}
where $P_0$ is again a constant of integration. Applying the boundary conditions on the inner and outer radius requires
\begin{align}
\mu\left(1 +\frac{4}{5}   R_1'(r_1)\right)+(P_f-P_1(r_1))\left(1+\frac{4 R_1(r_1)}{5 r_1}\right) &=0\\
\mu\left(1+\frac{4}{5}   R_1'(r_2)\right) -P_1(r_2) \left(1+\frac{4 R_1(r_2)}{5 r_2}\right)&=0.
\end{align}
We would like to solve these for the two constants of integration, $P_0$ and $c_4$, in terms of the applied pressure $P_f$. Unfortunately, the equations do not have an algebraic solution for $c_4$, but we can solve them for $P_f$ and $P_0$ in terms of $c_4$ to get
\begin{align}
P_0&=\mu  \left(\frac{5 r_2^2}{a^2} \left(1-\sqrt{\frac{c_4^2}{r_2^2}+1}\right)+\frac{5 c_4^2}{a^2} \log \left(\sqrt{c_4^2+r_2^2}+r_2\right)-\frac{c_4^2}{2 \left(c_4^2+r_2^2\right)}-\frac{1}{2} \log \left(\frac{c_4^2}{r_2^2}+1\right)+1\right)\\
P_f&=\frac{1}{2} \mu  \left(\frac{10 c_4^2}{a^2} \log \left(\frac{\sqrt{c_4^2+r_2^2}+r_2}{\sqrt{c_4^2+r_1^2}+r_1}\right)+10 \left(\frac{r_1^2}{a^2} \left(\sqrt{1+\left(\frac{c_4}{r_1}\right)^2}-1\right)-\frac{r_2^2}{a^2} \left(\sqrt{1+\left(\frac{c_4}{r_2}\right)^2}-1\right)\right)\right.\\
&\left.-\frac{r_1^2}{c_4^2+r_1^2}+\log \left(\frac{c_4^2}{r_1^2}+1\right)+\frac{r_2^2}{c_4^2+r_2^2}-\log \left(\frac{c_4^2}{r_2^2}+1\right)\right).\label{SI-Pf}
\end{align}
This fully specifies the base state. If we have a generic base state $\mathbf{u_1}$ and $P_1$ giving rise to effective deformation $G_1$, and we add small perturbations
\begin{equation}
\mathbf{u}=\mathbf{u_1}+\epsilon \mathbf{u_2}\hspace{3em} P=P_1+\epsilon P_2
\end{equation}
then expanding eqns.\ (\ref{SI-E_L1}-\ref{SI-bc}) about the base state to first order in $\epsilon$ yields
\begin{align}
&\frac{8\mu  }{ a^2 }\mathbf{u_2}=\frac{4\mu  }{5}\nabla^2 \mathbf{u_2}- \nabla P_2\cdot\mathrm{adj}\left(G_1\right)-\frac{4}{5}\nabla P_1\cdot \mathrm{adj}\left(\nabla \mathbf{u}_2\right)\label{SI-pert1}\\
&\Tr{\mathrm{adj}\left(G_1\right)\cdot\nabla\mathbf{u_2}}=0\label{SI-pert2}\\
&\left(\mu \frac{4}{5}\nabla \mathbf{u}_2-P_2\mathrm{adj}\left(G_1\right)^T+\frac{4}{5}\left(P_f-P_1\right)\mathrm{adj}\left(\nabla \mathbf{u_2}\right)^T\right)\cdot\mathbf{\hat{n}}=0.\label{SI-pertbc}
\end{align}
Where  $\mathrm{adj}$ denotes the adjugate matrix which, in 2-D, has the form
 \begin{equation}
\mathrm{adj}\left( \begin{array}{cc}
a&b\\
c&d
\end{array}\right)=\Det{\begin{array}{cc}
a&b\\
c&d
\end{array}}\left( \begin{array}{cc}
a&b\\
c&d
\end{array}\right)^{-1}=\left( \begin{array}{cc}
d&-b\\
-c&a
\end{array}\right).
\end{equation}
We now return to our circular base-state, and  take explicitly oscillatory forms for the perturbations
\begin{align}
\mathbf{u}&=R_1(r)\mathbf{\hat{r}}+\epsilon (R_2(r)\cos(n \theta) \mathbf{\hat{r}}+\Theta_2(r)\sin(n \theta) \boldsymbol{\hat{\theta}})\\
P&=P_1(r)+\epsilon P_2(r) \cos(n \theta),
\end{align}
from which we get
\begin{equation}
\nabla \mathbf{u}_2=\left( \begin{array}{cc}
R_2'(r)\cos(n \theta)&-\frac{nR_2(r)+\Theta_2(r)}{r}\sin(n \theta)\\
\Theta_2'(r)\sin(n \theta)&\frac{R_2(r)+n\Theta_2(r)}{r}\cos(n \theta)
\end{array}\right),\hspace{3em}\nabla P_2=\left( \begin{array}{c}
P_2'(r)\cos(n \theta)\\
-n P_2(r)\sin(n \theta)
\end{array}\right)
\end{equation}
so, the $\theta$ component of eqn.\ (\ref{SI-pert1}) is an algebraic equation for $P_2(r)$ solved by
\begin{equation}
P_2(r)=\frac{4 \left(\Theta_2(r) \left(\mu  \left(a^2 \left(n^2+1\right)+10 r^2\right)+a^2 r P_1'(r)\right)+a^2 n R_2(r) \left(2 \mu +r P_1'(r)\right)-a^2 \mu  r \left(r \Theta_2''(r)+\Theta_2'(r)\right)\right)}{a^2 n r \left(4 R_1'(r)+5\right)}
\end{equation}
and similarly, since $G_1$ is diagonal (see eqn.\ \ref{SI-G1eqn}), eqn.\ \ref{SI-pert2} is an algebraic equation for $\Theta_2$ solved by
\begin{equation}
\Theta_2(r)=-\frac{(4 R_1(r)+5 r) R_2'(r)}{n \left(4 R_1'(r)+5\right)}-\frac{R_2(r)}{n}=-\frac{\left(c_4^2+r^2\right) R_2'(r)+r R_2(r)}{n r}.
\end{equation}
The $r$ component of eqn.\ (\ref{SI-pert1}) ode for $R_2(r)$:
\begin{align}
a^2& \left(4 R_2(r) \left(\mu +\mu  n^2+r P_1'(r)\right)+4 n\Theta_2(r) \left(2 \mu +r P_1'(r)\right)+r \left((4 R_1(r)+5 r) P_2'(r)-4 \mu  \left(r R_2''(r)+R_2'(r)\right)\right)\right)\notag\\&+40 \mu  r^2 R_2(r)=0.
\end{align}
After substituting in the above forms for $R_1$, $P_1$, $\Theta_2$ and $P_2$ this is a non-linear fourth order differential equation. It is accompanied by the first order corrections to the four boundary conditions in eqn.\ \ref{SI-bc}:
\begin{align}
4 (P_f-P_1(r_1)) (n \Theta_2(r_1)+R_2(r_1))-P_2(r_1) (4 R_1(r_1)+5 r_1)+4 \mu  r_1 R_2'(r_1)&=0\\
(P_f-P_1(r_1)) (n R_2(r_1)+\Theta_2(r_1))+\mu  r_1 \Theta_2'(r_1)&=0\\
4 P_1(r_2) (n \Theta_2(r_2)+R_2(r_2))+P_2(r_2) (4 R_1(r_2)+5 r_2)-4 \mu  r_2 R_2'(r_2)&=0\\
P_1(r_2) (n R_2(r_2)+\Theta_2(r_2))-\mu  r_2 \Theta_2'(r_2)&=0.
\end{align}
Even evaluating the full form of the above equations, by substituting in the known fields, results in very cumbersome expressions. Solving the system analytically is a hopeless task. However, they are straightforward to solve using the Matlab's bvp4c boundary value solver. We input the equations, specify values for $n$, $r_1$, $r_2$ and $a$ then bvp4c is able to find the lowest value of $c_4$ for which the equations have a solutions, and find the solution.  We then iterate over $n$ until we find the solution with the lowest value of $c_4$ (that is the lowest  displacement on the inner boundary), to find the first unstable mode, which sets the threshold and mode-number for fingering. We finally use eqn.\ (\ref{SI-Pf}) to recover the fluid pressure threshold from the value of $c_4$. The threshold and mode-number predictions from this procedure are shown in fig.\ \ref{numsol} in our main manuscript.

\end{document}